\documentclass[twocolumn,showpacs,amsmath,amssymb]{revtex4}
\usepackage{graphicx}
\usepackage{dcolumn}
\usepackage{bm}

\begin{document}
\preprint{APS/123-QED}
\title[Stability of frozen waves in the Modified Cahn--Hilliard model]{Stability of frozen waves in the Modified Cahn--Hilliard model}
\author{E.S. Benilov}
 \altaffiliation[]{MACSI, Department of Mathematics, University of Limerick, Ireland}
 \email{Eugene.Benilov@ul.ie}
 \homepage{http://www.staff.ul.ie/eugenebenilov/hpage/}
\author{W.T. Lee}
 \altaffiliation[]{MACSI, Department of Mathematics, University of Limerick, Ireland}
 \email{William.Lee@ul.ie}
 \homepage{http://www3.ul.ie/wlee/index.html}
\author{R.O. Sedakov}
 \altaffiliation[]{MACSI, Department of Mathematics, University of Limerick, Ireland}
 \email{Roman.Sedakov@ul.ie}
\date{\today}

\begin{abstract}
We examine the existence and stability of frozen waves in diblock
copolymers with local conservation of the order parameter, which
are described by the modified Cahn--Hilliard model. It is shown
that a range of stable waves exists and each can emerge from a
`general' initial condition (not only the one with the lowest
density of free energy). We discuss the implications of these
results for the use of block copolymers in templating
nanostructures.
\end{abstract}

 \maketitle

\section{Introduction}

The Cahn--Hilliard equation \cite{CarterTaylorCahn97} is often
used to model microstructures arising from spinodal decomposition
in, say, polymer mixtures. One of the simplest systems exhibiting
this behavior would be a mixture of two polymers made from
monomers, A and B, with distinct chemical properties -- e.g., if A
is hydrophilic whereas B is hydrophobic. In such cases, a monomer
unit is attracted to units of the same type, while being repelled
by the other type, implying that the most energetically favorable
state is the one where A and B units are fully segregated. Such a
tendency is indeed confirmed by numerical modelling of the
Cahn--Hilliard equation \cite{deMelloFilho05} and is also in
agreement with stability analysis of similar models
\cite{BenilovChugunova10}.

One feature of the Cahn--Hilliard model is that the order
parameter is conserved globally (reflecting the mass conservation
law). The standard model, however, can be modified for
microstructures where the order parameter is conserved
\emph{locally}\ \cite{OhtaKawasaki86}. The modified model applies,
for example, if chains of the A and B monomers are parts of the
same polymer molecule, known as a `block copolymer'
\cite{HadjichristidisPispasFloudas03}, in which case they can
never be separated by a distance larger than the size of a single
molecule.

Systems with locally conserved order parameter are of particular
interest in nanotechnology. In particular, block copolymers are
used to template nanopatterns at surfaces, on scales that are too
small for traditional top-down photolithography
\cite{FitzgeraldEtAl09}. Such patterns have to be `directed'\
using chemical pre-patterning or topography, which is known as
graphoepitaxy and can provide excellent pattern registry
\cite{StuenEtAl10}. In its simplest form, conventional
lithographic techniques are used to create trenches in a silicon
wafer -- then the trenches are filled with block copolymer which
orders into lamellae parallel to the sidewall on annealing
\cite{FarrellEtAl10}. Finally a selected monomer is chemically
etched away and the remaining polymer used as an etch mask to
facilitate pattern transfer to the substrate, creating nanowires
on a scale too fine to be manufactured by conventional techniques
\cite{BorahEtAl11,MorrisEtAl12}.

The lamellae used to template the nanowires correspond to
\emph{frozen waves} (i.e. periodic time-independent solutions) of
the one-dimensional version of the modified Cahn--Hilliard
equation. It is particularly important whether these solutions are
unique or perhaps multiple stable solutions exist, as the latter
would impede one's control over the manufacturing process.

The present paper answers the above question by examining the
existence and stability of frozen waves in the modified
Cahn--Hilliard equation. In Sect. \ref{Section 2}, we shall
formulate the problem mathematically. In Sect. \ref{Section 3},
the \emph{existence }of frozen-wave solutions will be discussed.
In Sect. \ref{Section 4}, we shall present the results of a
\emph{stability} analysis of frozen waves.

\section{Formulation\label{Section 2}}

Consider a one-dimensional diblock polymer, with the
characteristic thickness $l$ of the A/B interface and mobility $M$
(the latter characterizes the diffusion of the order parameter
$\phi$). Using $l$ and $l^{2}/M$ to non-dimensionalize the spatial
coordinate $x$ and time $t$ respectively, we can write the
one-dimensional version of the modified Cahn--Hilliard equation
(MCHE) in the form%
\begin{equation}
\phi_{t}+\left(  \phi-\phi^{3}+\phi_{xx}\right)  _{xx}+\alpha\phi=0, \label{1}%
\end{equation}
where $\alpha$ determines the ratio of the characteristic size of
the region over which the order parameter is conserved to $l$.

As shown in Ref. \cite{LiuGoldenfeld89}, the MCHE admits frozen waves only if%
\[
0\leq\alpha<\tfrac{1}{4},
\]
whereas the wavelength (spatial period) $\lambda$ must satisfy%
\begin{equation}
\frac{2\pi}{\sqrt{\tfrac{1}{2}\left(  1+\sqrt{1-4\alpha}\right)  }}%
<\lambda<\frac{2\pi}{\sqrt{\tfrac{1}{2}\left(  1-\sqrt{1-4\alpha}\right)  }%
}\label{2}%
\end{equation}
(see Fig. 1). Ref. \cite{LiuGoldenfeld89} also computed examples
of frozen waves and the energy density $E$ as a function of a
frozen wave's length $\lambda$ -- which turned out to have a
minimum at a certain $\lambda =\lambda_{0}$. Given that the energy
cannot grow and is, thus, a Lyapunov functional, a conclusion was
drawn that $\lambda_{0}$ is stable.

\begin{figure}
\includegraphics[width=83mm]{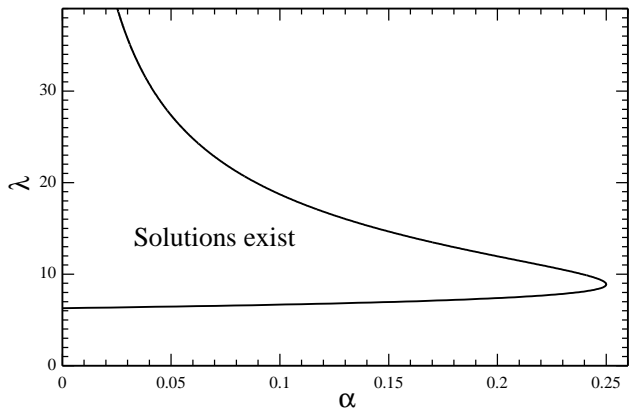}
\caption{The existence region of frozen waves (those described by
one-wave solutions) on the $\left(  \alpha,\lambda\right)  $-plane
($\alpha$ is the parameter in the modified Cahn--Hilliard equation
(\ref{1}), $\lambda$ is the wavelength). The boundaries of the
region are given by (\ref{2}).} \label{fig1}
\end{figure}

However, the fact that $\lambda_{0}$ minimizes the energy of
frozen waves means only that the corresponding wave is stable with
respect to perturbations \emph{of its length}, but not necessarily
to \emph{arbitrary} perturbations (for which the `general' second
variation of $E$ needs to be examined). On the other hand,
stability rarely occurs for a single value of a parameter --
hence, if $\lambda_{0}$ is indeed stable, it is likely to be one
of a range of stable wavelengths.

In what follows, we shall clarify the above issues by examining
the stability of frozen waves through the standard linear
analysis, not involving energy arguments.

To do so, we introduce frozen waves as time-independent solutions,
$\phi
=\bar{\phi}(x)$, for which Eq. (\ref{1}) yields%
\begin{equation}
\left(  \bar{\phi}-\bar{\phi}^{3}+\bar{\phi}_{xx}\right)
_{xx}+\alpha
\bar{\phi}=0. \label{3}%
\end{equation}
Together with the condition of spatial periodicity,%
\begin{equation}
\bar{\phi}(x+\lambda)=\bar{\phi}(x), \label{4}%
\end{equation}
Eq. (\ref{3}) determines $\bar{\phi}(x)$.

To examine the stability of a frozen wave $\bar{\phi}(x)$, assume%
\begin{equation}
\phi=\bar{\phi}(x)+\tilde{\phi}(t,x), \label{5}%
\end{equation}
where $\tilde{\phi}$ describes a small disturbance. Substituting
(\ref{5})
into Eq. (\ref{1}) and linearizing it, we obtain%
\begin{equation}
\tilde{\phi}_{t}+\left(
\tilde{\phi}-3\bar{\phi}^{2}\tilde{\phi}+\tilde{\phi
}_{xx}\right)  _{xx}+\alpha\tilde{\phi}=0. \label{6}%
\end{equation}
We confine ourselves to disturbances with exponential dependence
on $t$ (which
are usually a reliable indicator of stability in general),%
\begin{equation}
\tilde{\phi}(x,t)=\psi(x)\,\operatorname{e}^{st}, \label{7}%
\end{equation}
where $s$ is the disturbance's growth/decay rate. Substitution of
(\ref{7})
into (\ref{6}) yields%
\begin{equation}
s\psi+\left(  \psi-3\bar{\phi}^{2}\psi+\psi_{xx}\right)
_{xx}+\alpha\psi=0.
\label{8}%
\end{equation}
Unlike the base wave $\bar{\phi}$, the disturbance $\psi$ does not
have to be periodic; it is sufficient that the latter is bounded
at infinity. Given that $\psi$ is determined by an ordinary
differential equation with periodic coefficients [Eq. (\ref{8})],
the assumption of boundedness amounts to the
standard Floquet condition,%
\begin{equation}
\psi(x+\lambda)=\psi(x)\,\operatorname{e}^{\mathrm{i}\theta}, \label{9}%
\end{equation}
where $\theta$ is a real constant. Physically, condition (\ref{9})
implies that, if the disturbance propagates by one wavelength of
the base solution, the disturbance's amplitude remains the same,
whereas its phase may change by a value of $\theta$.

Eqs. (\ref{8})--(\ref{9}) form an eigenvalue problem, where $s$
and $\psi$ are the eigenvalue and the eigenfunction. If, for some
values of the phase shift $\theta$, one or more eigenvalues exist
such that $\operatorname{Re}s>0$, the corresponding base wave
$\bar{\phi}(x)$ is unstable.

\section{Frozen wave solutions\label{Section 3}}

It turns out that a lot of physically meaningful information about
frozen waves can be obtained in the limit of weak nonlinearity,
i.e. under the
assumption%
\[
\left\vert \bar{\phi}\right\vert \ll1.
\]
In order to understand qualitatively what to expect in this case,
one can simply omit the nonlinear term in Eq. (\ref{2}) and seek a
solution of the
resulting linear equation in the form%
\begin{equation}
\bar{\phi}=\varepsilon\cos\left(  kx+p\right)  , \label{10}%
\end{equation}
where the wave's amplitude $\varepsilon$ and phase $p$ are
arbitrary, whereas
the wavenumber $k$ satisfies%
\begin{equation}
-k^{2}\left(  1-k^{2}\right)  +\alpha=0. \label{11}%
\end{equation}
Assuming $k>0$ and recalling the relationship between the
wavelength and the
wavevector,%
\begin{equation}
\lambda=\frac{2\pi}{k}, \label{12}%
\end{equation}
one can see that, in the linear approximation, only two
wavelengths are allowed -- coincidentally, the same values which
represent the upper and lower boundaries of the existence region
(\ref{2}), shown in Fig. 1). Under the \emph{weakly nonlinear}
approximation, in turn, one should expect the wavelength to be
close, but not necessarily equal, to one of the above two values,
with a deviation from them proportional to some degree of the
wave's amplitude $\varepsilon$. Solutions similar to (\ref{10})
are the ones computed in Ref. \cite{LiuGoldenfeld89}; they will be
referred to as `one-wave solutions'.

To understand the physical meaning of \emph{two}-wave solutions,
seek a
solution of the linearized version of Eq. (\ref{3}) in the form%
\begin{equation}
\bar{\phi}=\varepsilon_{1}\cos\left(  k_{1}x+p_{1}\right)
+\varepsilon
_{2}\cos\left(  k_{2}x+p_{2}\right)  , \label{13}%
\end{equation}
where $\varepsilon_{1,2}$ and $p_{1,2}$ are arbitrary and
$k_{1,2}$ are the
roots of Eq. (\ref{11}) such that%
\begin{equation}
0<k_{1}<k_{2}. \label{14}%
\end{equation}
Physically, solution (\ref{13}) represents a superposition of two
periodic waves of zero frequency, with wavenumbers $k_{1,2}$ and
phases $p_{1,2}$. Note that (\ref{13}) is periodic only if
$k_{2}/k_{1}$ (and, hence, $\lambda _{2}/\lambda_{1}$) are both
rational numbers, which occurs only for some values of $\alpha$
(several examples of such are illustrated in Fig. 2).

\begin{figure}
\includegraphics[width=83mm]{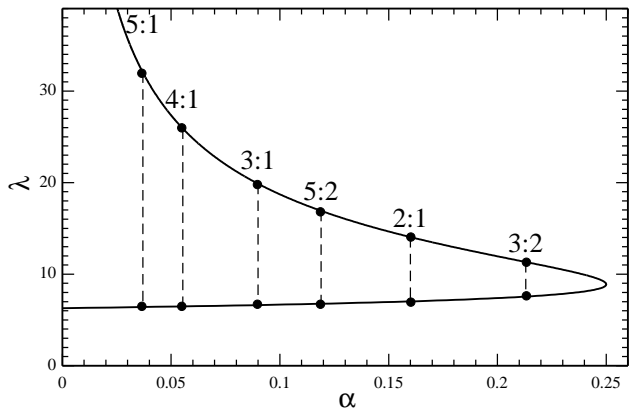}
\caption{A schematic illustrating linear two-wave solutions. The
solid curve [determined by (\ref{11})--(\ref{12})] corresponds to
linear frozen waves. The lengths $\lambda_{1,2}$ of the two
`composite' waves are shown by dots and are connected by dashed
lines. Two-wave solutions exist only for those values of $\alpha$
for which $\lambda_{2}/\lambda_{1}$ is a rational number
(presented in the figure above the corresponding $\lambda_{2}$).
The wavelength of a two-wave solution as a whole equals the lowest
common multiple of $\lambda _{1}$ and $\lambda_{2}$.} \label{fig2}
\end{figure}

The above linear analysis, however, leaves several important
questions unanswered. On the $\left(  \lambda,\alpha\right)
$-plane, for example, one-wave solutions seem to exist near curve
(\ref{11})--(\ref{12}), whereas Ref. \cite{LiuGoldenfeld89} found
frozen waves only \emph{inside} this curve, and not \emph{outside}
(see Fig. 1). This discrepancy -- as well as the question of
existence of two-wave solutions -- will be clarified in Sects.
\ref{Subsection 3.1}--\ref{Subsection 3.2}. The case of strong
nonlinearity will be examined numerically for both types of frozen
waves in Sect. \ref{Subsection 3.3}.

\subsection{Asymptotic results: one-wave solutions\label{Subsection 3.1}}

It is convenient to change the spatial coordinate $x$ to%
\[
\xi=kx,
\]
[where $k$ is, again, the wavenumber determined by (\ref{12})]. In
terms of
$\xi$, Eqs. (\ref{3})--(\ref{4}) become%
\begin{equation}
k^{2}\left(
\bar{\phi}-\bar{\phi}^{3}+k^{2}\bar{\phi}_{\xi\xi}\right)
_{\xi\xi}+\alpha\bar{\phi}=0, \label{15}%
\end{equation}%
\begin{equation}
\bar{\phi}(x+2\pi)=\bar{\phi}(x). \label{16}%
\end{equation}
We shall seek a solution as a series of the form%
\begin{equation}
\bar{\phi}=\varepsilon\left(  \bar{\phi}^{(0)}+\varepsilon^{2}\bar{\phi}%
^{(2)}+\cdots\right)  , \label{17}%
\end{equation}
and also expand the wavenumber $k$,%
\begin{equation}
k^{2}=K^{(0)}+\varepsilon^{2}K^{(2)}+\cdots. \label{18}%
\end{equation}
To leading order, Eqs. (\ref{15})--(\ref{16}) reduce to%
\begin{equation}
K^{(0)}\left(
\bar{\phi}^{(0)}+K^{(0)}\bar{\phi}_{\xi\xi}^{(0)}\right)
_{\xi\xi}+\alpha\bar{\phi}^{(0)}=0, \label{19}%
\end{equation}%
\[
\bar{\phi}^{(0)}(x+2\pi)=\bar{\phi}^{(0)}(x).
\]
We seek a solution in the form%
\[
\bar{\phi}^{(0)}=A\cos(n\xi+p),
\]
where $A$ and $p$ are real constants and $n>0$ is an integer. It
is sufficient to examine the case $n=1$, as $n\geq2$ corresponds
to re-defining the solution's spatial period by including more
than one wavelengths in it, without changing anything physically.
We also let $A=1$ (as the wave's physical amplitude still remains
arbitrary due to the arbitrariness of $\varepsilon$) and $p=0$ (as
a phase constant can always be included later).
Thus, the leading-order solution becomes%
\begin{equation}
\bar{\phi}^{(0)}=\cos\xi. \label{20}%
\end{equation}
Substitution of (\ref{20}) into (\ref{19}) yields%
\begin{equation}
K^{(0)}=\tfrac{1}{2}\left(  1\pm\sqrt{1-4\alpha}\right)  . \label{21}%
\end{equation}
In the next-to-leading order, Eq. (\ref{15}) yields%
\begin{multline*}
K^{(2)}\left(
\bar{\phi}^{(0)}+K^{(0)}\bar{\phi}_{\xi\xi}^{(0)}\right)
_{\xi\xi}\\
+K^{(0)}\left(
\bar{\phi}^{(2)}-\bar{\phi}^{(0)3}+K^{(2)}\bar{\phi}_{\xi\xi
}^{(0)}+K^{(0)}\bar{\phi}_{\xi\xi}^{(2)}\right)  _{\xi\xi}\\
+\alpha\bar{\phi}^{(2)}=0,
\end{multline*}
which, upon substitution of (\ref{20}), becomes%
\begin{multline}
K^{(0)}\left(
\bar{\phi}^{(2)}+K^{(0)}\bar{\phi}_{\xi\xi}^{(2)}\right)
_{\xi\xi}+\alpha\bar{\phi}^{(2)}\\
=\left(  K^{(2)}-2K^{(2)}K^{(0)}-\frac{3K^{(0)}}{4}\right)  \cos\xi\\
-\frac{9K^{(0)}}{4}\cos3x, \label{22}%
\end{multline}
This equation can have a $2\pi$-periodic solution only if the term
involving
$\cos\xi$ on the right-hand side vanishes, which implies%
\begin{equation}
K^{(2)}=\frac{3K^{(0)}}{4\left(  1-2K^{(0)}\right)  }, \label{23}%
\end{equation}
after which (\ref{22}) yields%
\begin{equation}
\bar{\phi}^{(2)}=\frac{9K^{(0)}}{36K^{(0)}\left(
1-9K^{(0)}\right)  -4\alpha
}\cos3\xi. \label{24}%
\end{equation}
Recalling that $K$ is related to the wavenumber $k$ by (\ref{18})
-- and $k$ is, in turn, related to the wavelength $\lambda$ by
(\ref{12}), one can use
expressions (\ref{21}) and (\ref{23}) to obtain%
\begin{equation}
\lambda_{1,2}=\frac{2\sqrt{2}\pi}{\sqrt{1\pm\sqrt{1-4\alpha}}}\left(
1\pm\frac{3\varepsilon^{2}}{8\sqrt{1-4\alpha}}\right)  +\operatorname{O}%
(\varepsilon^{4}), \label{25}%
\end{equation}
where $_{1}$ and $_{2}$ correspond to $+$\ and $-$\ respectively.
For $\varepsilon=0$ (i.e., in the linear limit), $\lambda_{1}$ and
$\lambda_{2}$ are the lower and upper boundaries of the existence
interval in Fig. 1, whereas an increase in $\varepsilon$ causes an
increase in $\lambda_{1}$ and a decrease in $\lambda_{2}$. This
means that one-wave solutions do not exist above $\left(
\lambda_{2}\right)  _{\varepsilon=0}$ and below $\left(
\lambda_{1}\right)  _{\varepsilon=0}$ -- which agrees with the
conclusions of Ref. \cite{LiuGoldenfeld89}.

Observe that expansion (\ref{25}) fails in the limit
$\alpha\rightarrow \frac{1}{4}$. This case should be examined
separately, by assuming
\begin{equation}
\alpha=\tfrac{1}{4}+\varepsilon^{2}\alpha^{(2)}. \label{26}%
\end{equation}
An expansion similar to the one above yields (technical details omitted)%
\begin{equation}
k^{2}=\tfrac{1}{2}\pm\sqrt{-\varepsilon^{2}\alpha^{(2)}-\tfrac{3}%
{8}\varepsilon^{2}}+\operatorname{O}(\varepsilon^{2}), \label{27}%
\end{equation}
from which $\lambda$ can be readily found. The corresponding
expansion of the
solution is (technical details omitted)%
\begin{equation}
\bar{\phi}=\varepsilon\left[
\cos\xi-\tfrac{9}{128}\varepsilon^{2}\cos
3\xi+\operatorname{O}(\varepsilon^{4})\right]  . \label{28}%
\end{equation}
Formulae (\ref{27})--(\ref{28}) describe the one-wave solution
near the `tip' ($\alpha\rightarrow\frac{1}{4}$,
$\lambda\rightarrow2\sqrt{2}\pi$) of the existence region shown in
Fig. 1. Formula (\ref{25}), in turn, describes the solution near
the boundary of the region, but not too close to its tip.

Finally, note that substitution of expression (\ref{21}) for
$K^{(0)}$ into
expression (\ref{24}) shows that $\bar{\phi}^{(2)}$ becomes infinite for%
\[
\alpha=\tfrac{9}{100},\hspace{0.4cm}\lambda=2\sqrt{10}\pi.
\]
This is another example where the straightforward expansion for
the one-wave
solution is inapplicable (in addition to the `tip point', $\alpha=\frac{1}{4}%
$, $\lambda=2\sqrt{2}\pi$). The unbounded growth of the
first-order term involving $\cos3\xi$ suggest that, in this case,
$\bar{\phi}^{(0)}$ should include both $\cos\xi$ and $\cos3\xi$ --
and not just the former term, as solution (\ref{20}) does.

This case will be examined in the next section.

\subsection{Asymptotic results: two-wave solutions\label{Subsection 3.2}}

As mentioned above, two-wave solutions exist near those values of
$\alpha$ for which the ratio of the wavenumbers of the two
individual waves is a rational number. Let $\alpha^{(0)}$ be one
of such values, with $\alpha^{(2)}$ being a
small deviation from it,%
\begin{equation}
\alpha=\alpha^{(0)}+\varepsilon^{2}\alpha^{(2)}. \label{29}%
\end{equation}
Substitution of this expression, together with expansions (\ref{17}%
)--(\ref{18}), in Eq. (\ref{15}) yields, to leading order,%
\begin{equation}
K^{(0)}\left(
\bar{\phi}^{(0)}+K^{(0)}\bar{\phi}_{\xi\xi}^{(0)}\right)
_{\xi\xi}+\alpha^{(0)}\bar{\phi}^{(0)}=0. \label{30}%
\end{equation}
We assume that the two waves of which a two-wave solution consists
have wavenumbers $K^{(0)}n_{1}$ and $K^{(0)}n_{2}$, where
$n_{1,2}$ are integers (without loss of generality, they can be
assumed to be coprime and such that $n_{1}>n_{2}>0$). Accordingly,
we shall seek a solution of Eq. (\ref{30}) in
the form%
\begin{equation}
\bar{\phi}^{(0)}=a\cos(n_{1}x+\theta)+\cos n_{2}x \label{31}%
\end{equation}
(the fact that the second wave's amplitude is unity and its phase
is zero does
not reduce generality). Substituting (\ref{31}) into (\ref{30}), we obtain%
\begin{equation}
\alpha^{(0)}=\frac{n_{1}^{2}n_{2}^{2}}{\left(
n_{1}^{2}+n_{2}^{2}\right)
^{2}},\hspace{0.4cm}K^{(0)}=\frac{1}{n_{1}^{2}+n_{2}^{2}}. \label{32}%
\end{equation}
Recalling that $K$ is related to the wavelength $\lambda$, one can
see that these equalities determine the points in the $\left(
\alpha,\lambda\right) $-plane near which two-wave weakly-nonlinear
solutions are localized (they are illustrated in Fig. 2).

To the next-to-leading order, (\ref{15}) yields\begin{widetext}%
\begin{multline*}
K^{(0)}\left(
\bar{\phi}^{(2)}+K^{(0)}\bar{\phi}_{\xi\xi}^{(2)}\right)
_{\xi\xi}+\alpha^{(0)}\bar{\phi}^{(2)}=K^{(0)}\left(
\frac{3a}{2}\cos (n_{1}x+\theta)+\frac{a^{3}}{4}\left[
3\cos(n_{1}x+\theta)+\cos
3(n_{1}x+\theta)\right]  \right.  \\
+\frac{3a^{2}}{4}\left\{  \cos\left[  \left(  2n_{1}-n_{2}\right)
x+2\theta\right]  +\cos\left[  \left(  2n_{1}+n_{2}\right)
x+2\theta\right] \right\}  +\frac{3a}{4}\left\{  \cos\left[
\left(  n_{1}-2n_{2}\right) x+\theta\right]  +\cos\left[  \left(
n_{1}+2n_{2}\right)  x+\theta\right]
\right\}  \\
+\left.  \frac{3a^{2}}{2}\cos n_{2}x+\frac{1}{4}\left(  3\cos n_{2}%
x+\cos3n_{2}x\right)  \right)  _{\xi\xi}-a\left[
K^{(2)}n_{1}^{2}\left(
2K^{(0)}n_{1}^{2}-1\right)  +\alpha^{(2)}\right]  \cos(n_{1}x+\theta)\\
-\left[  K^{(2)}n_{2}^{2}\left(  2K^{(0)}n_{2}^{2}-1\right)
+\alpha ^{(2)}\right]  \cos n_{2}x.
\end{multline*}
\end{widetext}This equation has a periodic solution for $\bar{\phi}^{(2)}$
only if the terms involving $\cos(n_{1}x+\theta)$ and $\cos
n_{2}x$ cancel out, but the specifics depend on whether $\left(
n_{1},n_{2}\right)  =\left( 3,1\right)  $ or not.

In the former case, straightforward calculations yield\begin{widetext}%
\[
K^{(2)}=-\dfrac{3\left(  7a^{3}-a^{2}+17a+3\right)
}{340a},\hspace
{0.4cm}\alpha^{(2)}=-\dfrac{9\left(  9a^{3}+3a^{2}+9a+1\right)  }{420a}%
\hspace{0.4cm}\text{if}\hspace{0.4cm}\left(  n_{1},n_{2}\right)
=\left( 3,1\right)  ,
\]
whereas the latter case yields%
\[
K^{(2)}=-\dfrac{3\left[  \left(  n_{1}^{2}-2n_{2}^{2}\right)  a^{2}+2n_{1}%
^{2}-n_{2}^{2}\right]  }{4\left(  n_{1}^{2}+n_{2}^{2}\right)
\left(
n_{1}^{2}-n_{2}^{2}\right)  },\hspace{0.4cm}\alpha^{(2)}=-\dfrac{9n_{1}%
^{2}n_{2}^{2}\left(  a^{2}+1\right)  }{4\left(
n_{1}^{2}+n_{2}^{2}\right)
^{2}}\hspace{0.4cm}\text{if}\hspace{0.4cm}n_{1}\neq3n_{2}.
\]
\end{widetext}Now, using expressions (\ref{17})--(\ref{18}), (\ref{29}),
(\ref{12}), (\ref{31})--(\ref{32}) to relate $\bar{\phi}^{(2)}$,
$K^{(2)}$, and $\alpha^{(2)}$ to the `physical' quantities, we can
summarize the two-wave
solution in the form%
\begin{equation}
\bar{\phi}=\varepsilon\left[  a\cos(n_{1}x+\theta)+\cos
n_{2}x\right]
+\operatorname{O}(\varepsilon^{3}), \label{33}%
\end{equation}
and\begin{widetext}%
\begin{equation}
\left.
\begin{tabular}
[c]{l}%
$\alpha=\dfrac{9}{100}-\varepsilon^{2}\dfrac{9\left(  9a^{3}+3a^{2}%
+9a+1\right)  }{420a}+\operatorname{O}(\varepsilon^{4}),\medskip$\\
$\lambda=2\sqrt{10}\pi+\varepsilon^{2}\dfrac{3\pi\sqrt{10}\left(  7a^{3}%
-a^{2}+17a+3\right)  }{34a}+\operatorname{O}(\varepsilon^{4}),$%
\end{tabular}
\right\}  \hspace{0.4cm}\text{if}\hspace{0.4cm}\left(
n_{1},n_{2}\right)
=\left(  3,1\right)  ,\label{34}%
\end{equation}
or%
\begin{equation}
\left.
\begin{tabular}
[c]{l}%
$\alpha=\dfrac{n_{1}^{2}n_{2}^{2}}{\left(  n_{1}^{2}+n_{2}^{2}\right)  ^{2}%
}-\varepsilon^{2}\dfrac{9n_{1}^{2}n_{2}^{2}\left(  a^{2}+1\right)
}{4\left( n_{1}^{2}+n_{2}^{2}\right)
^{2}}+\operatorname{O}(\varepsilon^{4}),\medskip
$\\
$\lambda=2\pi\sqrt{n_{1}^{2}+n_{2}^{2}}+\varepsilon^{2}\dfrac{3\pi\left[
\left(  n_{1}^{2}-2n_{2}^{2}\right)
a^{2}+2n_{1}^{2}-n_{2}^{2}\right]
\sqrt{n_{1}^{2}+n_{2}^{2}}}{4\left(  n_{1}^{2}-n_{2}^{2}\right)
}+\operatorname{O}(\varepsilon^{4}),$%
\end{tabular}
\right\}  \hspace{0.4cm}\text{if}\hspace{0.4cm}n_{1}\neq3n_{2}.\label{35}%
\end{equation}
\end{widetext}Expressions (\ref{34}) and (\ref{35}) can be viewed as
parametric representations (with $\varepsilon$ and $a$ being the
parameters) of the existence region of two-wave solutions with
$n_{1,2}$, on the $\left( \alpha,\lambda\right)  $-plane. It can
be shown that, for all $n_{1,2}$, this region is a `semi-infinite
sector' (see an example in Fig. 3). However, since we assumed weak
nonlinearity, this conclusion can only be trusted near the
vertices of the sectors. Effectively, we have found the tangent
lines to the boundaries of the `true' region of the existence
interval.

\begin{figure}
\includegraphics[width=83mm]{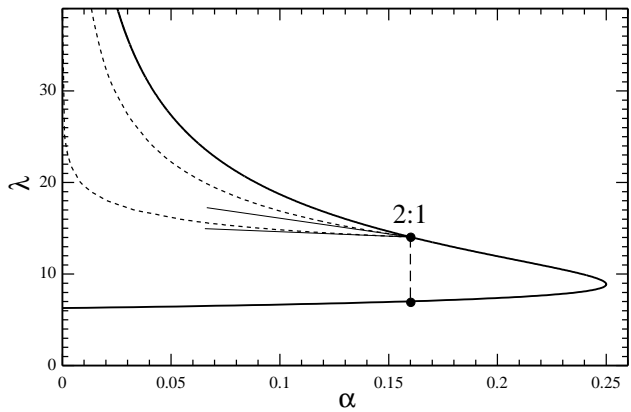}
\caption{Regions of existence (bounded by the dotted line) of
two-wave solutions with $\lambda _{2}/\lambda_{1}=2$\/$:$\/$1$.
Thin solid lines show the boundaries of the existence region
calculated under the assumption of weak nonlinearity. The thick
solid line shows the existence region for one-wave solutions (as
in Fig. 1).} \label{fig3}
\end{figure}

Observe that $\varepsilon^{2}$ can be eliminated from (\ref{34})
and
(\ref{35}), which yields\begin{widetext}%
\begin{equation}
\dfrac{\alpha-\tfrac{9}{100}}{\lambda-2\sqrt{10}\pi}=-\dfrac{6\left(
9a^{3}+3a^{2}+9a+1\right)  }{27\sqrt{10}\pi\left(
7a^{3}-a^{2}+17a+3\right)
}+\operatorname{O}(\varepsilon^{2}),\hspace{0.4cm}\text{if}\hspace
{0.4cm}\left(  n_{1},n_{2}\right)  =\left(  3,1\right)  ,\label{36}%
\end{equation}
or%
\begin{equation}
\frac{\alpha-\dfrac{n_{1}^{2}n_{2}^{2}}{\left(
n_{1}^{2}+n_{2}^{2}\right)
^{2}}}{\lambda-2\pi\sqrt{n_{1}^{2}+n_{2}^{2}}}=-\dfrac{9n_{1}^{2}n_{2}%
^{2}\left(  a^{2}+1\right)  \left(  n_{1}^{2}-n_{2}^{2}\right)
}{3\pi\left( n_{1}^{2}+n_{2}^{2}\right)  ^{5/2}\left[  \left(
n_{1}^{2}-2n_{2}^{2}\right)
a^{2}+2n_{1}^{2}-n_{2}^{2}\right]  }+\operatorname{O}(\varepsilon^{2}%
)\hspace{0.4cm}\text{if}\hspace{0.4cm}n_{1}\neq3n_{2}.\label{37}%
\end{equation}
\end{widetext}For given $n_{1,2}$ and $\left(  \alpha,\lambda\right)  $,
(\ref{36}) and (\ref{37}) can be treated as equations for $a$ [it
is, essentially, the ratio of the amplitudes of the waves which
constitute the two-wave solution -- see (\ref{33})]. Observe that,
(\ref{37}) admits two roots for $a$ (with equal magnitudes and
opposite signs), which means that two-wave solutions with
$n_{1}\neq3n_{2}$ exist in pairs. In some cases these solutions
can be obtained from each other by shifting\ $x\rightarrow
x+\operatorname{const}$ (for the 2:1 case, for example, $\operatorname{const}%
=\pi$) -- but in other cases, the solutions with positive and
negative $a$ seem to be genuinely different.

For the case $\left(  n_{1},n_{2}\right)  =\left(  3,1\right)  $,
in turn, it follows from (\ref{37}) that three roots exist for
$a$. One of the three, however, corresponds to the \emph{one}-wave
solution. Indeed, recall that the expansion derived for those
failed near $\alpha=\tfrac{9}{100}$, $\lambda=2\sqrt{10}\pi$, i.e.
precisely where the asymptotic theory for \emph{two}-wave
solutions predicts existence of those with $\left(
n_{1},n_{2}\right)  =\left(  3,1\right)  $. In fact, one- and
two-wave solutions cannot be distinguished in this region, as, in
both cases, the coefficients of $\cos3x$ and $\cos x$ are of the
same order.

\subsection{Numerical results\label{Subsection 3.3}}

In this section, we shall present examples of strongly-nonlinear
one- and two-wave solutions and the existence region for the
latter. Two numerical methods have been used: the shooting method
(which turned out to be insufficient for large wavelengths) and
the method of Newton relaxation (which worked marginally better).

Figs. 4a and 4b show examples of increasingly nonlinear one- and
two-wave solutions respectively, for a fixed wavelength $\lambda$
and decreasing $\alpha$. In Fig. 4a, observe the increase of the
wave's amplitude as $\alpha$ is moving\ away from the boundary of
the existence region (for this value of $\lambda$, the boundary is
located at $\alpha\approx0.131$). Fig. 4b, in turn, illustrates
the fact that the margins of the existence region for two-wave
solutions correspond to the cases where the amplitude of one of
the two waves vanishes (which is how they bifurcate from one-wave
solutions).

\begin{figure}
\includegraphics[width=83mm]{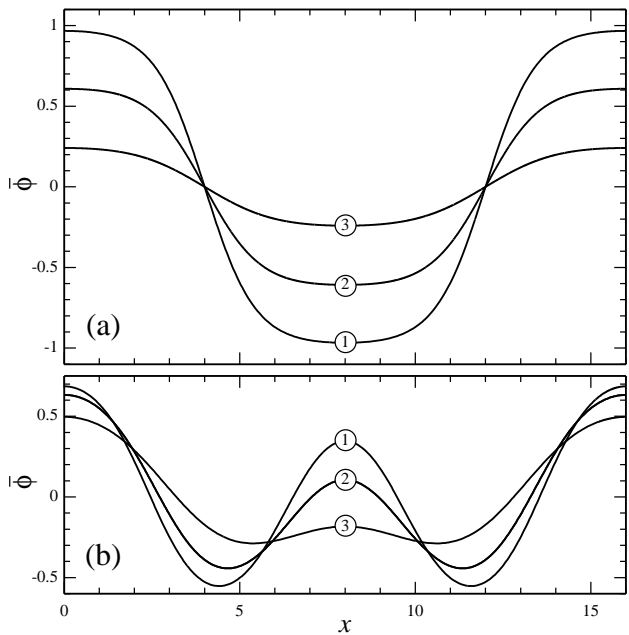}
\caption{Examples of frozen waves with spatial period
$\lambda=16$: (a) one-wave solutions with
$\alpha=0.005,~0.050,~0.100$ (curves 1, 2, 3 respectively); (b)
two-wave solutions with $\alpha=0.07,~0.09,~0.11$ (curves 1, 2, 3
respectively).} \label{fig4}
\end{figure}

\section{The stability of frozen waves\label{Section 4}}

\subsection{Asymptotic results}

We shall first examine the stability of frozen waves
asymptotically, under the same assumption of weak nonlinearity
used to find the frozen waves themselves.

We shall start by re-writing the stability problem
(\ref{8})--(\ref{9}) in
terms of $\xi=kx$, which yields%
\begin{equation}
s\psi+k^{2}\left(
\psi-3\bar{\phi}^{2}\psi+k^{2}\psi_{\xi\xi}\right)
_{\xi\xi}+\alpha\psi=0, \label{38}%
\end{equation}%
\begin{equation}
\psi(\xi+2\pi)=\psi(\xi)\,\operatorname{e}^{\mathrm{i}\theta}. \label{39}%
\end{equation}
It turns out that all weakly nonlinear solutions, one- and
two-wave alike, are unstable for all $\alpha$ except
$\alpha\rightarrow\frac{1}{4}$ (i.e. except the `tip' of the
existence interval for one-wave solutions shown in Fig. 1). We
emphasize that this conclusion does not apply to \emph{strongly}
nonlinear waves (which can be either stable or unstable for any
$\alpha$ -- see the next section).

Thus, assuming that
$\alpha=\frac{1}{4}+\operatorname{O}(\varepsilon^{2})$, we
represent $\alpha$ by expression (\ref{26}); the corresponding
base solution, $\bar{\phi}$, is represented by
(\ref{27})--(\ref{28}).

A straightforward analysis shows that, if the phase shift $\theta$
is order-one, all solutions of the eigenvalue problem
(\ref{38})--(\ref{39}), with the parameters determined by
(\ref{26})--(\ref{28}), are stable. Thus, instability can occur
only for small $\theta$, which can be conveniently
accounted for by letting%
\begin{equation}
\theta=2\pi\varepsilon q, \label{40}%
\end{equation}
where $q$ is an order-one constant. Note that the smallness of the
phase shift $\theta$ implies that the instability occurs at
wavelengths close to that of the base solution.

It is also convenient to introduce%
\begin{equation}
\psi_{new}=\psi\operatorname{e}^{-\mathrm{i}\varepsilon qx}. \label{41}%
\end{equation}
Using (\ref{40})--(\ref{41}) to replace $\theta$ and $\psi$ in
Eqs. (\ref{38})--(\ref{39}) with $q$ and $\psi_{new}$, we obtain
(subscript
$_{new}$ omitted):%
\begin{multline}
s\psi+k^{2}\left(  \psi+k^{2}\psi_{\xi\xi}-3\varepsilon^{2}\phi^{2}%
\psi\right)  _{\xi\xi}\\
+\mathrm{i}\varepsilon qk^{2}\left(
2\psi_{\xi}+4k^{2}\psi_{\xi\xi\xi }\right)
-\varepsilon^{2}q^{2}k^{2}\left(  \psi+6k^{2}\psi\right)  _{\xi\xi
}+\operatorname{O}(\varepsilon^{3})\\
+\alpha\psi=0, \label{42}%
\end{multline}%
\begin{equation}
\psi(\xi+2\pi)=\psi(\xi), \label{43}%
\end{equation}
where the specific form of the terms
$\operatorname{O}(\varepsilon^{3})$ will not be needed.

It can be demonstrated that no instability occurs if
$s=\operatorname{O}(1)$ -- hence, only small $s$ need to be
examined. We assume (and shall eventually justify by obtaining a
consistent asymptotic expansion) that
\[
s=\varepsilon^{2}s^{(2)}+\cdots,
\]
while the eigenfunction is%
\[
\psi=\psi^{(0)}+\varepsilon^{2}\psi^{(2)}+\cdots.
\]
To leading order, Eqs. (\ref{42})--(\ref{43}),
(\ref{26})--(\ref{28}) reduce
to%
\[
\tfrac{1}{2}\left(
\psi^{(0)}+\tfrac{1}{2}\psi_{\xi\xi}^{(0)}\right)
_{\xi\xi}+\tfrac{1}{4}\psi^{(0)}=0,
\]%
\[
\psi^{(0)}(\xi+2\pi)=\psi^{(0)}(\xi),
\]
which yield
\begin{equation}
\psi^{(0)}=S\sin x+C\cos x, \label{46}%
\end{equation}
where $S$ and $C$ are undetermined constants.

In the next-to-leading order, Eqs. (\ref{42})--(\ref{43}), (\ref{26}%
)--(\ref{28}) yield (after straightforward algebra)\begin{widetext}%
\begin{multline*}
\tfrac{1}{2}\left(  \psi_{\xi\xi}^{(2)}+\tfrac{1}{2}\psi_{\xi\xi\xi\xi}%
^{(2)}\right)  +\tfrac{1}{4}\psi^{(2)}+\tfrac{3}{4}\left[  S\sin
x+C\cos x+\tfrac{1}{2}S\left(  3\sin3x-\sin x\right)
+\tfrac{1}{2}C\left(
3\cos3x+\cos x\right)  \right]  \\
+\left(  s^{(2)}+q^{2}-\tfrac{3}{8}\right)  \left(  S\sin x+C\cos
x\right) -2\mathrm{i}q\sqrt{-\alpha^{(2)}-\tfrac{3}{8}}\left(
S\cos x-C\sin x\right) =0,
\end{multline*}%
\[
\psi^{(2)}(\xi+2\pi)=\psi^{(2)}(\xi).
\]
\end{widetext}This boundary-value problem has a solution for $\psi^{(2)}$ only
if%
\[
\tfrac{3}{8}S+\left(  s^{(2)}+q^{2}-\tfrac{3}{8}\right)  S+2\mathrm{i}%
q\sqrt{-\alpha^{(2)}-\tfrac{3}{8}}C=0,
\]%
\[
\tfrac{9}{8}C+\left(  s^{(2)}+q^{2}-\tfrac{3}{8}\right)  C-2\mathrm{i}%
q\sqrt{-\alpha^{(2)}-\tfrac{3}{8}}S=0,
\]
which, in turn, has a solution for $S$ and $C$ only if%
\[
s^{(2)2}+\left(  2q^{2}+\tfrac{3}{4}\right)  s^{(2)}+q^{2}\left[  q^{2}%
+\tfrac{3}{4}-4\left(  -\alpha^{(2)}-\tfrac{3}{8}\right)  \right]
=0.
\]
This equation determines the eigenvalue $s^{(2)}$. It can be
readily shown that $s^{(2)}$ is stable (i.e.
$\operatorname{Re}s^{(2)}<0$) for all $q$ only
if%
\[
\alpha^{(2)}<-\tfrac{9}{16}.
\]
Finally, using (\ref{26})--(\ref{27}) to express $\alpha^{(2)}$ in
terms of the `physical' parameters $\alpha$ and $\lambda$, we
obtain the following stability criterion for a frozen wave with
parameters $\left(  \alpha
,\lambda\right)  $:%
\begin{equation}
\left(  \lambda-2\sqrt{2}\pi\right)
^{2}\lesssim\tfrac{8}{3}\pi^{2}\left(
\tfrac{1}{4}-\alpha\right)  . \label{47}%
\end{equation}
Observe that, for any $\alpha<\frac{1}{4}$, there exists an
\emph{interval} of stable wavelengths, not just a single value of
$\lambda$.

As mentioned before, condition (\ref{47}) applies only if $\alpha$
is close to $\frac{1}{4}$. In the next subsection, it will be
extended numerically to arbitrary values of $\alpha$.

\subsection{Numerical results}

The eigenvalue problem (\ref{8})--(\ref{9}) was solved numerically
for $\psi(x)$ and $s$ with the base wave $\bar{\phi}(x)$ computed
using problem (\ref{3})--(\ref{4}).

The general features of the dispersion relation [the dependence
$s(\theta)$] of the eigenvalue problem (\ref{8})--(\ref{9}) is
described in the Appendix, whereas here we shall only present the
stability diagram on the $\left( \alpha,\lambda\right)  $ plane --
see in Fig. 5. Evidently, for all values of $\alpha$, an interval
of $\lambda$ exists where one-wave solution are stable -- which
confirms and extends the asymptotic
($\alpha\rightarrow\frac{1}{4}$) stability criterion (\ref{47}).
We have not found any stable two-wave solutions, which suggests
they are either unstable or perhaps their regions of stability are
small and difficult to locate.

\begin{figure}
\includegraphics[width=83mm]{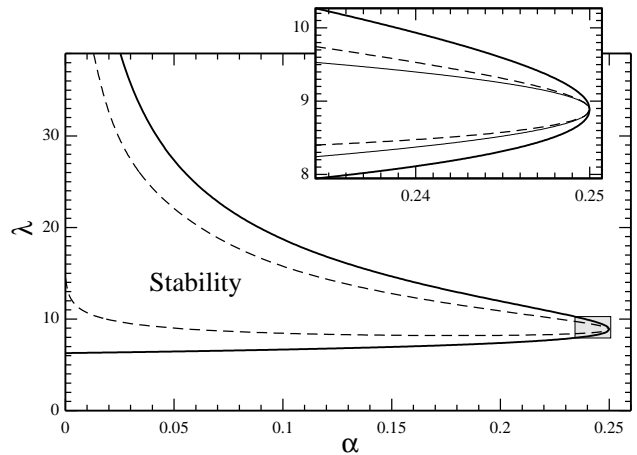}
\caption{The stability region of one-wave solutions on the $\left(
\alpha,\lambda\right)  $-plane (shown by the dashed line). The
thick solid line shows the existence region of one-wave solutions
(as in Fig. 1). The upper panel represents a blow-up of the shaded
region of the lower panel [the thin solid line shows the stability
region's asymptotic boundary, (\ref{47})].} \label{fig5}
\end{figure}

To illustrate that \emph{any} solution from the range of stable
frozen waves (not necessarily the wave with the minimum
free-energy density) can emerge from a `general' initial
condition, we have carried out the following numerical experiment.
The time-dependent MCHE (\ref{1}) was simulated using finite
differences with a fully implicit backwards Euler method, and the
results of the simulations presented below are for $\alpha=0.1$.

In this case, the energy minimizing wavelength is
$\lambda_{0}=11.31$ (calculated from the interpolation formula of
Ref. \cite{LiuGoldenfeld89}), with the corresponding frozen-wave
solution denoted by $\phi_{0}\left( x\right)  $. According to our
analysis, however, a solution $\phi_{1}\left( x\right)  $ with the
commensurate wavelength $\lambda_{1}=14.13$ ($\lambda
_{0}/\lambda_{1}=4/5$) should also be stable.

To verify this, Eq. (\ref{1}) was simulated in a domain of size
$L=5\lambda _{0}=4\lambda_{1}$ which accommodates both solutions.
The initial condition was chosen as a `mixture' of the frozen
waves $\phi_{0}\left(  x\right)  $ and $\phi_{1}\left(  x\right)
$, i.e.
\begin{equation}
\phi\left(  x,0\right)  =\beta\,\phi_{0}\left(  x\right)  +\left(
1-\beta\right)  \,\phi_{1}\left(  x\right)  ,\label{48}%
\end{equation}
where $\beta\in\left(  0,1\right)  $ is the `mixing ratio'. The
timestep was $0.1$ and $400$ gridpoints per period were used, and
it has been verified that the results were mesh and timestep
independent.

If $\phi_{0}(x)$ was the only stable solution, the system would
evolve towards $\phi_{0}$ for all $\beta$. Our simulations
nevertheless show that, for $\beta=0.2$, the system evolves back
to $\phi_{1}$ (see Fig. 6), which confirms our conclusion about
the existence of \emph{multiple} stable states.

\begin{figure}
\includegraphics[width=83mm]{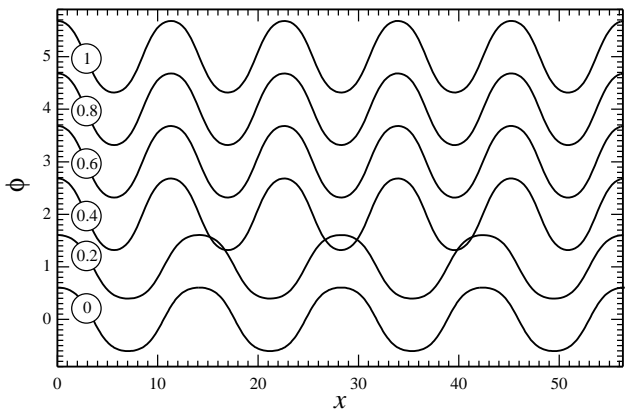}
\caption{Evolution of perturbed steady states within the stable
region at $\alpha=0.1$. The traces show the limiting
($t\rightarrow\infty$) solution initialized by (\ref{48}) for
various $\beta$. The traces are incremented in steps of $0.5$,
beginning with zero increment for $\beta=0$ (the lowest trace) and
ending for $\beta=1$ (the highest trace). The curves are marked
with the corresponding values of $\beta$.} \label{fig6}
\end{figure}

\section{Concluding remarks}

The main result of the present paper is illustrated in Fig. 5,
which shows the stability region of (one-wave) frozen solutions of
the modified Cahn--Hilliard equation (\ref{1}). We have also found
a new class of frozen waves -- the `two-wave solutions', but these
seem to be unstable and, thus, less important than the usual,
one-wave type.

We have also made a more general -- and potentially more important
-- conclusion regarding the energy approach to studies of
stability. If a family of solutions exists and one of them
minimizes the energy functional, this does not necessarily mean
that all other solutions are unstable. Furthermore, the stability
of the minimizer solution cannot be guaranteed either: even though
it is stable with respect to the perturbation of `shifting along
the family of solutions', another perturbation can still
destabilize it.

Physically, our results imply that when lamellar microstructures
of block copolymers are used to template nanowires, one must
ensure that only the desired state is created. This may become
more of a critical concern as larger numbers of nanowires are to
be created within a single trench. In practice, some control over
this can be exerted via the annealing schedule. It should also be
noted that a kinetically stable quenched state may be selected
rather than a true time independent solution to the modified
Cahn--Hilliard equation.

Finally, it would be interesting to extend the present results to
steady states with \emph{two} spatial dimensions, similar to those
found in Ref. \cite{ChoksiMarasWilliams11} for an equation similar
to the two-dimensional MCHE (but with a slightly different
nonlinearity).

\begin{acknowledgments}
The authors acknowledges the support of the Science Foundation
Ireland (RFP Grant 08/RFP/MTH1476 and Mathematics Initiative Grant
06/MI/005)
\end{acknowledgments}

\appendix{}

\section{The structure of the dispersion relation and the eigenfunctions of
the eigenvalue problem (\ref{8})--(\ref{9})}

First of all, it can be readily shown (and has been confirmed
numerically) that the eigenvalue $s$ of problem
(\ref{8})--(\ref{9}) is real.

\begin{figure}
\includegraphics[width=83mm]{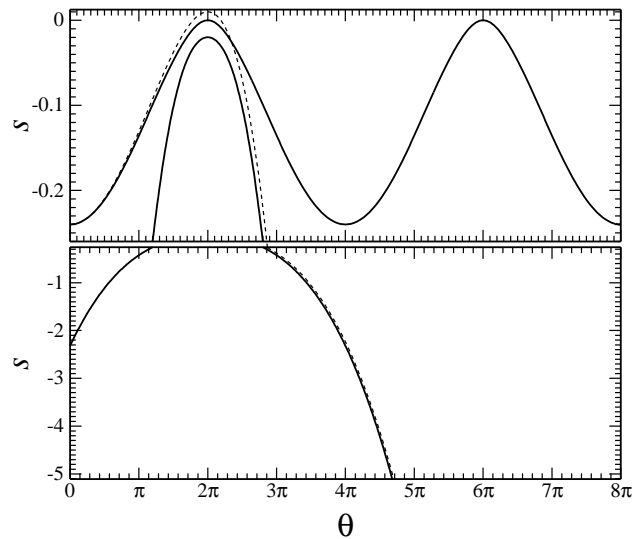}
\caption{The dispersion relation $s(\theta)$ for the eigenvalue
problem (\ref{8})--(\ref{9}) with $\alpha=0.24$, $\lambda=8.886$
(solid line). The dotted line shows the limiting dispersion curve
(\ref{A.2}). Note the difference in the vertical axes' scales of
the upper and lower panels.} \label{fig7}
\end{figure}

Observe also that, for the boundary points of the existence region
(see Fig. 1), frozen waves have zero amplitude, i.e.
$\bar{\phi}=0$. In this case, the
eigenfunction of problem (\ref{8})--(\ref{9}) can be readily found,%
\begin{equation}
\psi=\exp\frac{\mathrm{i}\theta x}{\lambda}, \label{A.1}%
\end{equation}
as well as the dispersion relation (i.e. the dependence of the
eigenvalue $s$
on the phase shift $\theta$),%
\begin{equation}
s=-\alpha+\left(  \frac{\theta}{\lambda}\right)  ^{2}-\left(
\frac{\theta
}{\lambda}\right)  ^{4}. \label{A.2}%
\end{equation}
Observe that, as follows from (\ref{A.1}), the eigenfunction
becomes more and more oscillatory with increasing $\theta$.

In the general case, (i.e. for \emph{interior} points of the
existence region of frozen waves, where $\bar{\phi}\neq0$), one
would expect that (\ref{A.2}) is somehow perturbed, but still
keeps its structure as a single curve on the $\left(
\theta,s\right)  $ plane. Furthermore, the large-$\theta$ part of
the `general' dispersion relation should not differ much from that
of (\ref{A.2}) -- as, for rapidly oscillating $\psi$, the term
involving $\bar{\phi}$ in equation (\ref{8}) is negligible.

Our computations show, however, that, no matter how small
$\bar{\phi}$ is, the dispersion relation splits into two branches
-- see Fig. 7.

The following properties of the two branches have been observed:

\begin{itemize}
\item The upper branch is periodic with a period of $4\pi$.

\item Both branches are symmetric with respect to $\theta=2\pi$
(provided they are extended to negative $\theta$).
\end{itemize}

\begin{figure}
\includegraphics[width=83mm]{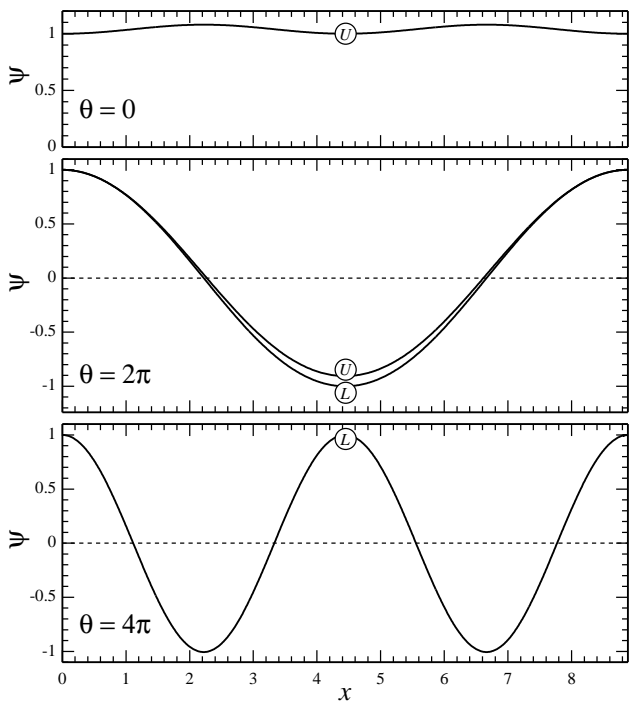}
\caption{Examples of eigenfunctions of problem
(\ref{8})--(\ref{9}) with $\alpha=0.24$, $\lambda=8.886$. The
eigenfunctions marked with \textquotedblleft U\textquotedblright\
and \textquotedblleft L\textquotedblright\ correspond to the upper
and lower branches, respectively, of the dispersion relation
illustrated in Fig. 6.} \label{fig8}
\end{figure}

As a result, the only `original' part of the upper branch is the
segment $\theta\in\left[  0,2\pi\right]  $, whereas the `original'
part of the lower branch is that for $\theta=\left[
2\pi,\infty\right)  $. Not surprisingly, the unperturbed
dispersion relation (\ref{A.2}) `switches' from the upper branch
to the lower one near the point $\theta=2\pi$ (see Fig. 7).

Fig. 8 shows typical behavior of the eigenfunctions: for
$\theta=0$, $\psi(x)$ does not oscillate at all; for
$\theta=2\pi$, it oscillates once; for $\theta=4\pi$, it
oscillates twice, etc.

\begin{figure}[b]
\includegraphics[width=83mm]{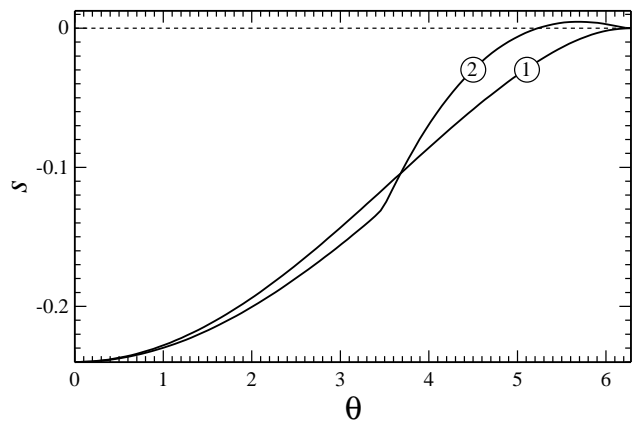}
\caption{The dispersion curves (upper branches) for $\alpha=0.24$
and (1) $\lambda=8.886$ (stability for all $\theta$), (2)
$\lambda=9.786$ (instability for sufficiently small $\theta$).}
\label{fig9}
\end{figure}

Fig. 9, in turn, shows the onset of instability brought by a
change of the period of the base wave. One can see that the waves
with $\theta\approx2\pi$ are first to lose stability (which agrees
with our asymptotic analysis of the case
$\alpha\rightarrow\frac{1}{4}$).

Finally, we mention that two exact solutions were found for the
eigenvalue
problem (\ref{8})--(\ref{9}):%
\begin{align}
s  &  =-\alpha\qquad\text{for}\qquad\theta=0,\label{A.3}\\
s  &  =0\hspace{1cm}\text{for}\qquad\theta=2\pi. \label{A.4}%
\end{align}
In the latter case the disturbance can be found analytically,
$\psi=\bar{\phi }$, and it corresponds to infinitesimal shift of
the base wave. The former solution does not seem to have an
obvious physical meaning (nor does it admit an obvious analytical
expression for the eigenfunction, as equality (\ref{A.4}) has been
established numerically).

\end{document}